# Controlling the growth of 2D conjugated coordination polymers to induce metallic and spin-dependent transport signatures

Hio-Ieng Un[1,2*], Jordi Ferrer Orri[1,3], Ian E. Jacobs[1], Naoya Fukui[4], Hiroshi Nishihara[4], Caterina Ducati[3], Samuel D. Stranks[1], Henning Sirringhaus[1*]

[1] Optoelectronics Group, Cavendish Laboratory, University of Cambridge, JJ Thomson Avenue, Cambridge CB3 0US, UK
[2] State Key Laboratory of Luminescent Materials and Devices, School of Materials Science and Engineering, South China University of Technology, Guangzhou 510641, China
[3] Department of Materials Science and Metallurgy, University of Cambridge, 27 Charles Babbage Road, Cambridge CB3 0FS, UK
[4] Research Institute for Science and Technology, Tokyo University of Science, Noda-shi, Chiba 278-8510, Japan
[*] Corresponding authors: hioiengun@scut.edu.cn (H.-I.U.); hs220@cam.ac.uk (H.S.)

## Abstract

Understanding growth evolution and thereby implementing precise microstructural tuning of two-dimensional (2D) conjugated coordination polymers (cCPs) is crucial to achieve efficient electronic conduction towards their full potential and to observe materials' intrinsic properties. However, fundamental understanding of how 2D cCPs films grow remains very limited. Here, we use copper-benzenehexathiol (Cu-BHT) cCP as a model system to unravel the growth evolution of layered films in liquid-liquid interfacial synthesis in order to identify strategies to achieve tuning of structure-property relationships. We find that thin films formed at the early stage of growth in 20 minutes facilitate smoother, denser, and horizontally oriented films, and thereby achieve higher electrical conductivity of > 3000 S cm$^{-1}$ with a metallic temperature dependence down to 20 K. They also reveal signatures of quantum interference mediated weak anti-localisation and Kondo-like effect in magnetotransport at low temperatures. These phenomena are not observed when long reaction time was employed. Our findings offer a new perspective for the growth of dynamically reversible self-assemblies, that is different from the traditional paradigm of longer reaction time being associated with higher ordering and performance, and offer a platform to study spin-related transport properties of these materials with higher performance for advanced electronic, thermoelectric, and potential spintronic applications.

# Introduction

Two-dimensional (2D) conjugated metal-organic frameworks (cMOFs) or conjugated coordination polymers (cCPs) are an emerging class of functional materials for which the electronic and spintronic properties can be programmed through tailored design and precise synthesis.[1–8] Compared to one-dimensional conjugated polymers, their 2D conjugation framework provides additional transport pathways to bypass structural and energetic defects and suppresses vibrational degrees of freedom, enabling greater defect-tolerance and faster charge transport, while their topological structures offer potentially a versatile platform to explore intriguing quantum properties, arising in 2D Kagome lattices comprising magnetic ions with interesting spin textures.[9] However, at present this potential remains not fully realized experimentally due to the randomly oriented, polycrystalline nature of their thin films comprising a high density of defects. This limits the achievable electrical conductivities and also prevents the observation of more intrinsic transport phenomena that reflect their quantum electronic structure and spin textures. Fundamental understanding of their growth mechanism and strategies to precisely control their structure at various length scales remain a prerequisite for exploring their intrinsic and efficient electronic and quantum properties.

The main challenge is that the very different growth dynamics between the in-plane coordination bonds and the cross-plane van der Waals and π-π interactions makes the growth often faster in the cross-plane than the in-plane direction; this causes the crystals or aggregates to grow and become orientated along the less delocalized layer-stacked direction.[10–12] Traditional approaches involving varying solvent, concentration, and reaction temperature are capable of realizing high crystallinity but usually need days to complete the self-correction of reversible chemical bonds and interactions. More importantly, these traditional approaches have failed in controlling anisotropic, orientated growth.[13–16] Novel effective ways to control orientation and morphology involve using complex methods[17–20] such as templates and surfactants; however, use of templates is only applicable to limited cases and use of surfactants significantly reduce electrical conduction.[17,21] Finding template-free, facile, rapid, and yet well adjustable synthetic approaches to realizing highly-ordered uniaxially orientated 2D cCP thin films with efficient electronic conduction seems a challenging trade-off.

In this work, we use the most electrically conductive copper-benzenehexathiol (Cu-BHT) cCP as a model system, to systematically investigate the morphological and structural changes of the films at the nano-/micro-scale at different growth stages in liquid-liquid interfacial synthesis (LLIS). Better understanding of the growth evolution enables good control of morphology and high level of uniaxial grain orientation during film synthesis by simply controlling reaction time. Uniaxial anisotropy of the layer stacking is surprisingly not obtained by longer reaction time of > 2 hours, typically favored for self-correction in self-assembled systems, but is favoured by shorter reaction time of < 20 minutes corresponding to the early stage of growth. In the rapidly prepared thin films with smooth, dense, and uniaxially anisotropic microstructure, metallic charge transport with higher electrical conductivity of > 3000 S cm$^{-1}$ can be achieved and quantum coherent, and spin-dependent transport signatures can be observed. Our work provides evidence that longer synthesis time is not always better for achieving more efficient electronic and quantum properties, offering new insights into controlling the aggregation and growth of dynamically reversible self-assembly driven by coordination and π–π interactions.

## Results and discussion

### More efficient electronic conduction in thinner films prepared by shorter reaction time

The LLIS used here is similar to our recently reported method[22] but with much shorter reaction time. Films with different defect levels were obtained by using constant BHT concentration and varying the $Cu^{2+}$ concentration, i.e. varying the Cu/BHT molar ratio in the reaction solutions (refer to as Cu/BHT ratio in this work). After gently adding certain amount of aqueous solution of $Cu(OAc)_2$ onto the upper part of the water phase and allowing the $Cu^{2+}$ ions to diffuse to the interface, nucleation is triggered, followed by growth, from top to down (**Figure 1a**), due to the hydrophobic nature. The difference from our previous work[22], where reaction time of 2 – 2.5 h was used, is that here we use reaction time as short as 15 – 20 min to obtain a film thickness < 100 nm (see **Figure S1** for the relation of film thickness with reaction time). We refer to the former as regular films and the latter as thinner films in this work.

These films show a clear negative correlation between film thickness ($d$) and electrical conductivity ($\sigma$) for both Cu/BHT ratios of 2 and 5 (**Figure 1b,c**, see **Figures S2, S3** for thickness determination and the thickness dependence of the electrical conductivity of other ratios). In the thinner films of Cu/BHT ratio of 5, conductivity of up to 3406 S cm$^{-1}$ can be achieved (**Figure 1c**). This is the highest electrical conductivity at room temperature achieved so far for 2D cCPs. The conductivity values are seemingly linear with inverse thickness ($1/d$) for the regular films (**Figure 1d**). Since at a given channel length and width, $\sigma \propto \frac{1}{R} \cdot \frac{1}{d}$; the approximately linear relation between $\sigma$ and $1/d$ means the resistances of the films are approximately constant, seemingly indicating that electrical conduction is not homogeneous across the bulk of the films and the most efficient conduction occurs in the portion of the films grown at early times (**Figure 1e**).

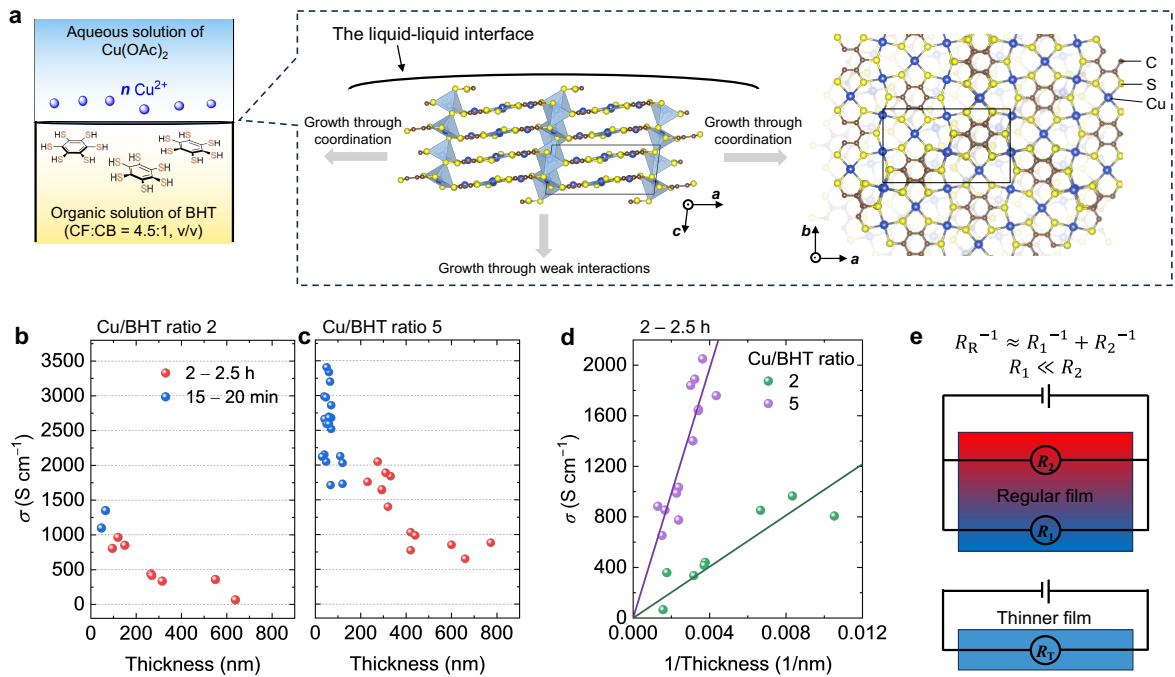

**Figure 1. Comparison of the electrical properties of Cu-BHT films prepared by 2 – 2.5 h and 15 – 20 min**. **a**, Illustration of the liquid-liquid interfacial synthesis (LLIS) and the molecular structure of Cu-BHT. *n* in the left of the diagram represents the molar ratio of the precursors $Cu^{2+}$ to BHT added into the reaction solutions, varying from 2 to 7 around the nominally ideal ratio of 3 for perfect $Cu_3BHT$ structure. **b,** Electrical conductivities at Cu/BHT ratio of 2 as a function of thickness. **c**, Electrical conductivities at Cu/BHT ratio of 5 as a function of thickness. The variation of the values is most likely due to the cracks and wrinkles, which are more easily present as well as more significantly influence the device-scale electrical conductivities in thin films of < 100 nm. **d**, Plot of electrical conductivity against inverse thickness for the films prepared by 2 – 2.5 h, showing an approximately proportional relation. **e**, Diagram of the electrical conduction in regular films and thinner films and their simplified circuits.

## Growth evolution and morphological control

To understand the correlation between reaction time, thickness, and electronic conduction, we monitored the surface and the bulk morphologies of the films as a function of Cu/BHT ratio and reaction time. Our key observations can be summarized as follows:

i. Among different compositions, a compact layer which is facing up to the aqueous solution in the synthesis can be observed, varying from approximately 20 to 200 nm in thickness as Cu/BHT ratio increases from 2 to 5 (**Figure 2a-c**). The higher the Cu/BHT ratio the thicker the compact layer becomes.

ii. The pH value of the aqueous phase (measured by indicator paper) changes from 10 to 5.5 after 2-hour reaction.

iii. The elementary distribution at the cross section of the films characterized by energy-dispersive X-ray spectroscopy (EDX) reveals that the Cu-to-S atomic ratio of the surface facing to the aqueous solution is closest to the ideal value of 0.5 and gradually deviates more towards the surface facing to the organic solution (**Figure 2d,e,f**), indicating that the nanosheets growing first and closer to the interface are more perfect.

These phenomena can be interpreted by a self-templated, diffusion- and thermodynamically-limited (ST-D&TL) synthesis and growth hypothesis: Since ideal Cu-BHT has an intrinsically dense, nonporous molecular structure, the layers that are already formed at the interface sufficiently hinder the mass (reactant) transport. As Cu/BHT ratio increases, the films change from more ideal to more defective (i.e. more ligand vacancies), similar to what was demonstrated in our previous work[22] (**Figure 2d,e,f**). There also appear to be more defect-induced nano-pores in the compact layer that allow mass transfer/diffusion to a position farther from the interface. This explains why the thickness of the compact layer increases as Cu/BHT ratio increases from 2 to 5 (phenomenon i), and also implies that the growth after the compact layer is presumably influenced by mass diffusion. Considering the hydrophobic and top-down growth nature of the films, solvated $Cu^{2+}$ and $AcO^-$ or neutral molecule $Cu(OAc)_2$ are the reactants which have to diffuse into the other phase through the compact layer for the reaction. The gradually reducing amount of reactants and the limited diffusion through the compact layer, and the gradually increasing concentration of $H^+$ (i.e. reducing pH value, as evident by phenomenon ii) over the reaction gradually shift the thermodynamic equilibrium away from the formation of deprotonated or coordinated BHT and thus limit the available amount of deprotonated or coordinated BHT that can take part in the later coordination reaction with $Cu^{2+}$ (**Scheme S1**). Consequently, the films became more BHT deficient and more Cu rich as they grow continuously from the interface towards the organic phase (**Figure 2d,e,f**) (phenomenon iii). The deprotonated BHT molecules tend to pre-assemble in a face-to-face orientation on the 2D conjugated plane of the Cu-BHT nanosheets at defective regions at which aggregation is usually more thermodynamically favorable. Therefore, vertically oriented downward growth along cross plane direction using the compact layer as self-template dominates the growth (**Figure 2m,n,o**). Such a molecular stacking orientation is further supported by diffraction experiments (see discussion below). A transformation of more laterally expanded, platelet-like surface morphology is observed as reducing BHT concentration and/or increasing $Cu^{2+}$ concentration, i.e. increasing Cu/BHT ratio from 2 to 5 (**Figure 2g-i, j-l**). All these results demonstrate that there is room to tune the film morphology by controlling the reactant ratio and the synthesis time. Increasing $Cu^{2+}$ concentration and reducing synthesis time favor producing a smoother and more compact morphology.

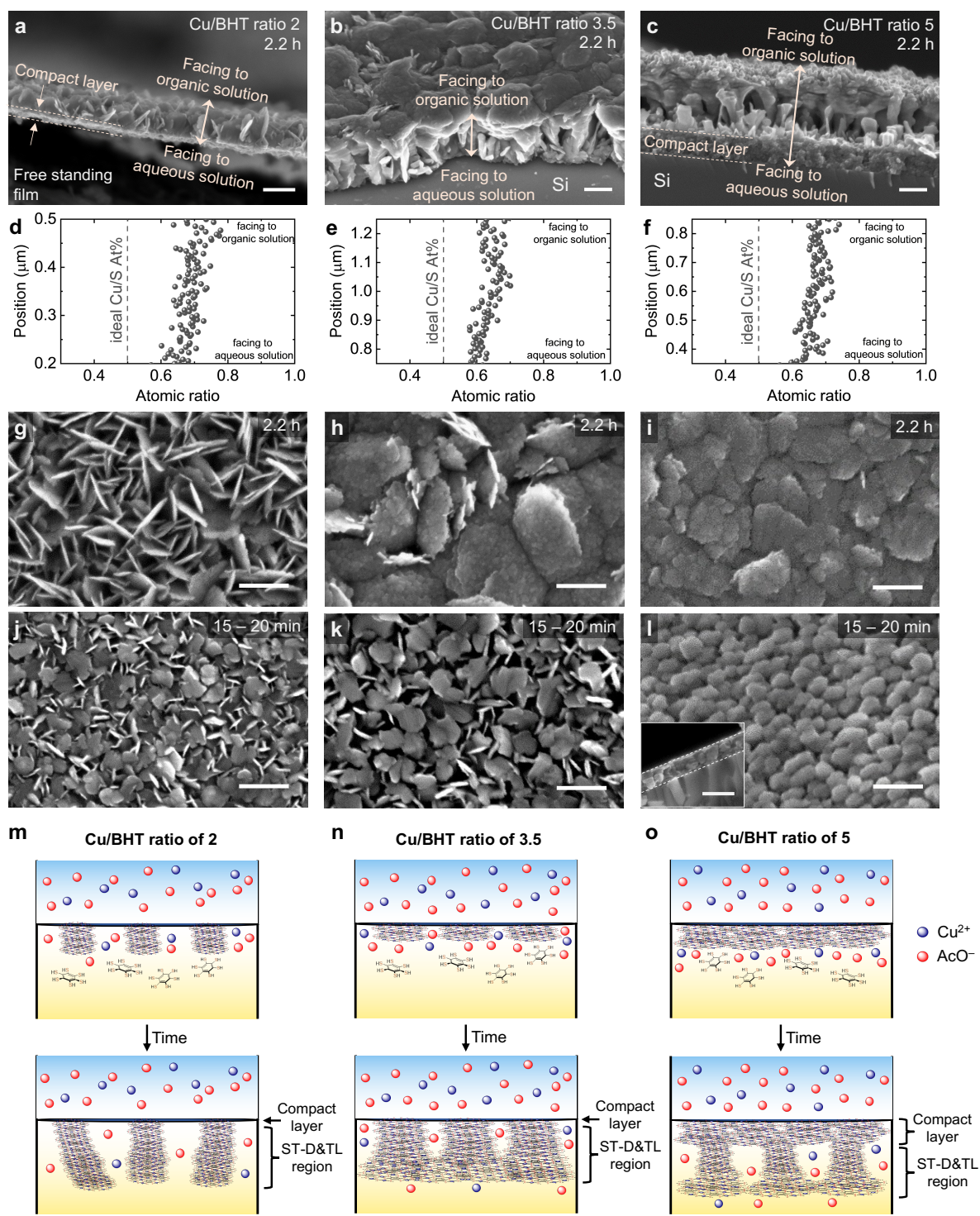

**Figure 2. SEM images and diagram for the growth evolution hypothesis of Cu-BHT films in liquid-liquid interfacial synthesis**. **a – c**, Cross section images of the regular films. **d – f**, Vertical distribution of Cu/S atomic ratio of the regular films. Corresponding cross sections are labelled in **a – c** using the double, line arrow. **g – i**, The surface facing the organic solution of the regular films with reaction time of 2.2 h. **j – l**, The surface facing the organic solution of the thinner films synthesized with 15 – 20 min reaction time. All scale bars here are 200 nm. **m – o**, Proposed growth situations at the beginning and at the mid-stage. From the left to the right column the images represent Cu-BHT ratios of 2, 3.5, and 5 respectively. ST-D&TL region stands for self-template, diffusion- and thermodynamic-limiting region.

**Characterization of structural order and anisotropy**

Scanning electron diffraction (SED) provides insight into spatially-resolved structural ordering at the nanoscale. Experiments were performed on the thinner films with a thickness of 50 – 100 nm, otherwise the electrons are difficult to transmit. The diffraction patterns were found to match with the non-van der Waals layered Cu$_3$BHT,[23] showing diffraction peak (002) found at 0.275 ± 0.015 Å$^{-1}$ (orange box) and diffraction peak (200) found at 0.126 ± 0.015 Å$^{-1}$ (blue box) for Cu/BHT ratios of 2 and 3.5 (**Figure 3b,c**). Since virtual dark field (vDF) mode only allows scattered electrons to transmit, the brighter the regions in vDF images correspond to the higher amount of the scattered electrons and thereby either the higher the crystallinity or the lower the local thickness. **Figure 3b,c** reveals that the orientation of the crystalline grains for Cu/BHT ratios of 2 and 3.5 are heterogeneous on the nanoscale; the particularly highly crystalline regions (blue boxes labelled as region 2) are usually face-on grains, suggesting that the in-plane lattice has a better structural ordering than the cross-plane lattice, as expected by the weaker interaction in the cross-plane than in the in-plane direction. In the films of Cu/BHT ratio of 5 the bright regions just represent the thinner local thickness that allows more electrons to transmit since the corresponding diffractions are rather weak and diffuse (**Figure 3d**) and grains are only crystalline at small length scale (**Figure S4**). This is consistent with our previous results that the higher the Cu/BHT ratio the poorer the crystallinity,[22] this is observed for both regular and thinner films.

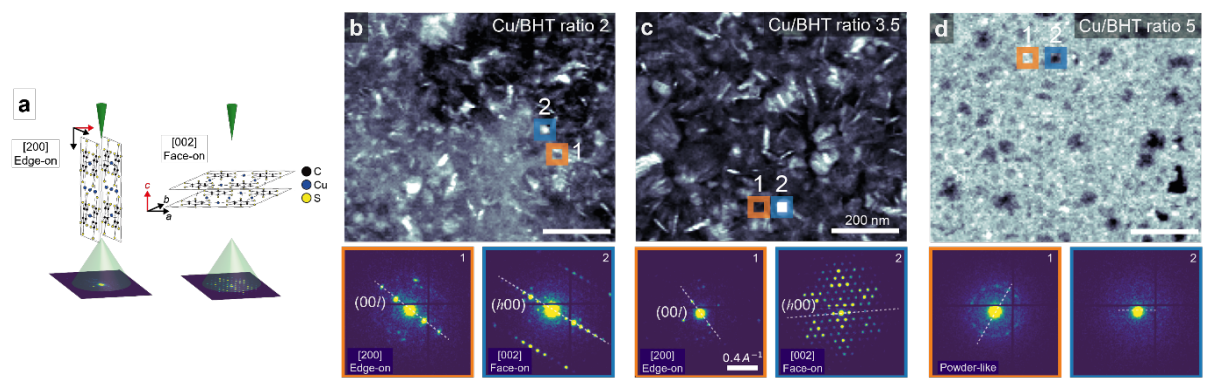

**Figure 3. SED characterization on the thinner films (50-100 nm) unraveling crystallinity and grain orientation spatially on the nanoscale**. **a**, Diagram of SED measurement where edge-on and face-on grains can be distinguished through the pattern and spacing of the electron diffraction in reciprocal space. **b – d**, Virtual dark field (vDF) images for films of Cu/BHT ratios of 2, 3.5, and 5 respectively. The vDF images are calculated by using the counts of the scattered electrons at each position in the scan area. The bottom orange and blue boxes exhibit the diffractions corresponding to regions 1 and 2 labeled in the vDF images. Labels 1 and 2 are for edge-on and face-on orientations, respectively. The brighter the regions in the vDF images the more scattered electrons to transmit.

Inspired by the cross-sectional elemental distribution of the films that shows lower chemical imperfection in the layers growing first (facing to the aqueous solution) compared to the layers growing afterward (facing to the organic solution) (**Figure 2d-f**), we investigated whether thinner films overall have better structural ordering than thicker regular films. We compared the structural ordering between the thicker, regular and the thinner films by grazing incidence wide-angle X-ray scattering (GIWAXS).

Thinner films of Cu/BHT ratios of 2 and 3.5 show lattice parameters closer to the reported crystal structure, with *d*-spacings found at $d_{(200)}$ = 7.33±0.018, 7.36±0.010 Å and $d_{(002)}$ = 3.39±0.001, 3.40±0.001 Å, with respect to the reported values of $d_{(200)}$ = 7.404 Å and $d_{(002)}$ = 3.418 Å found in crystals,[23] while thicker regular films produce $d_{(200)}$ = 7.28±0.015, 7.31±0.012 Å and $d_{(002)}$ = 3.38±0.001, 3.39±0.001 Å, respectively (**Figure 4i**). Interestingly, however, thinner films of Cu/BHT ratio of 5 show lattice parameters that deviate more from the crystal structure compared to the regular films, with $d_{(200)}$ = 7.26±0.017 Å and $d_{(002)}$ = 3.48±0.002 Å found for the thinner films and $d_{(200)}$ = 7.32±0.010 Å and $d_{(002)}$ = 3.41±0.002 Å for regular films (**Figure 4i**). Full width at half maximum (FWHM) of the diffraction peaks, and hence paracrystallinity is challenging to be quantitatively compared within the experimental errors due to limited signal-to-noise ratio in most cases, but we estimated that thinner films of Cu/BHT ratio of 5 have larger in-plane paracrystallinity and smaller in-plane X-ray coherence length than the regular films (**Figure 4j**). These results reveal that for films of Cu/BHT ratios of 2 and 3.5, shorter reaction time of 15 – 20 min gives raise to lattice parameters closer to those of single crystals with limited observable influence on crystalline quality. For a Cu/BHT ratio of 5, thin films grown with shorter reaction time appear less crystalline than those grown with longer reaction time, suggesting that degree of crystallinity is not the main factor responsible for the observed high conductivity.

Another key difference between the regular and the thinner films prepared by different reaction times is the level of uniaxial orientation. Although all the films have a mainly face-on stacking orientation, as evidenced by the diffraction peak (002) being mainly located in the $q_z$ direction, regular films synthesized with longer reaction time show a not very strong but detectable diffraction (002) signal in the in-plane $q_{xy}$ direction in general regardless of the Cu/BHT ratio (**Figure 4h**). This is also evident in the azimuthal map of the (002) diffraction peak for Cu/BHT ratio of 2 in **Figure 4k**, which indicates a weaker face-on anisotropy in the thicker films. Our structural analysis therefore suggests that the higher level of uniaxial orientation observed in the thinner films is probably the main reason for the improvement in their electronic properties.

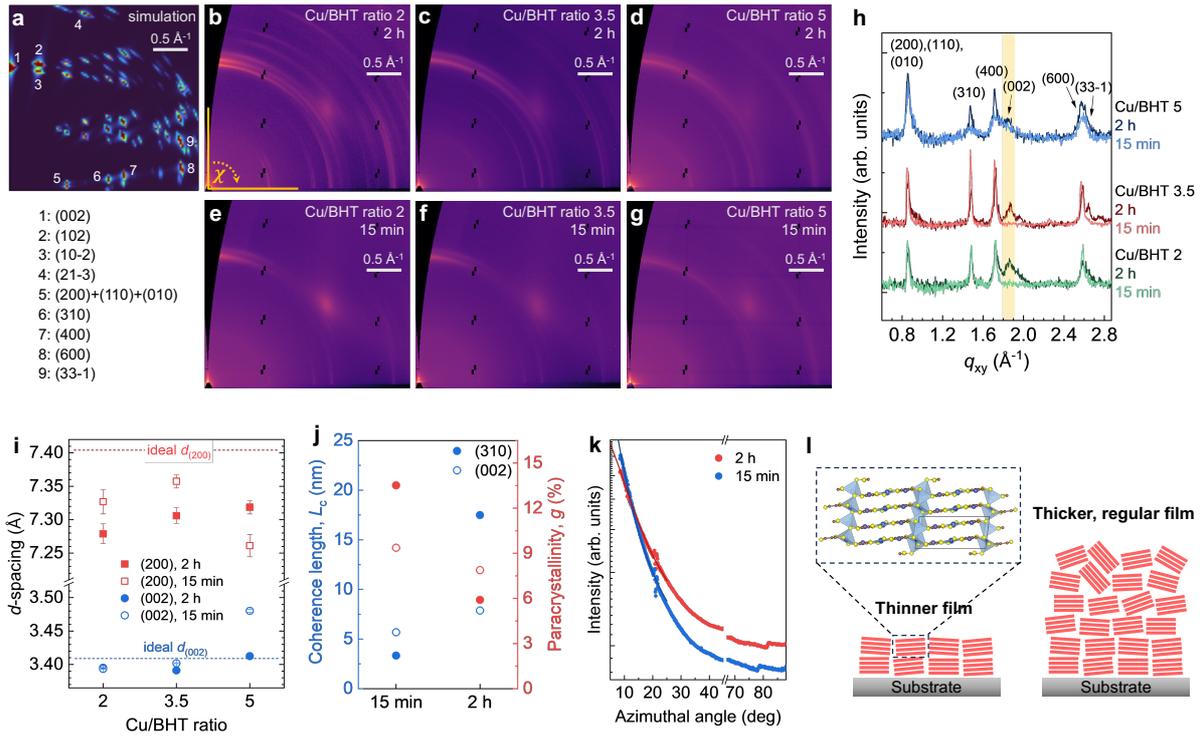

**Figure 4. GIWAXS characterization of the regular and the thinner films of Cu-BHT. a**, Simulated GIWAXS image of Cu$_3$BHT. **b-d**, GIWAXS images for the regular films of Cu/BHT ratios of 2, 3.5, and 5 synthesized by 2 h. **e-g**, GIWAXS images for the films of Cu/BHT ratios of 2, 3.5, and 5 synthesized by 15 min. **h**, In-plane diffraction profiles extracted from GIWAXS images. **i**, $d$-spacing extracted from diffraction peaks (200) and (002) with respect to their ideal values resolved from crystals (red and blue dash line). **j**, In-plane and out-of-plane coherence length and paracrystallinity of the films of Cu/BHT ratio 5. **k**, Azimuthal intensity profiles of (002) diffraction peak extracted from the GIWAXS images of Cu/BHT ratio 2 with the Ewald sphere correction. **l**, Diagram for the uniaxial orientation.

## Enhanced metallic and spin-dependent transport in thinner films

Temperature-dependent electrical conductivity and magnetoresistance (MR) measurements were performed to better understand the influence of the morphology and structural ordering on transport properties. Here MR is defined as $[\rho(B) - \rho(0)]/\rho(0)$, where $\rho$ is resistivity and $B$ is magnetic flux density. Temperature-dependent electrical conductivities reveal that with reducing film thickness, films of Cu/BHT ratio of 5 retain their metallic nature, i.e., their conductivities rise with decreasing temperature ($d\sigma/dT < 0$), and display a two-fold enhanced conductivity compared to the thicker films. Films of Cu/BHT ratio of 2 exhibit an approximately 3-fold enhanced conductivity compared to the thicker films and become "more metallic", i.e. they approach the regime with $d\sigma/dT < 0$, (**Figure 5a**). This can be interpreted as evidence that the more compact, smoother thinner films are less affected by grain boundary-limited transport (**Figure 2g,i,j,l**) and also benefit from the higher level of uniaxial face-on orientation (**Figure 4h,k,l**).

In MR measurements, with the magnetic field perpendicular to the sample plane and the current, the MR in the regular films of Cu/BHT ratios of both 2 and 5 as well as in the thinner films of Cu/BHT ratio of 2 (**Figure 5c**) is positive at temperature down to 9 – 10 K and exhibits a parabolic dependence on the magnetic field, as a result consistently reflecting their metallic nature. Interestingly, however, in the most highly conducting thin films of Cu/BHT ratio of 5 we observe a competition between a weak positive MR in the lower field region ($|B| < 2.5$ T) and a strong negative MR in the higher field region ($|B| > 2.5$ T) at the same temperature (**Figure 5d**). A positive, cusp-like feature in MRs in low field region is a typical signature of weak antilocalization (WAL) due to spin–orbit coupling (SOC), which is often observed in materials with heavy elements and originates from destructive phase interference of the electron wavefunction between time reversed trajectories affecting the probability for carrier backscattering.[24–27] Negative MRs can be due to electron-electron interaction (EEI) in diffusive transport regime,[28,29] spin-dependent scattering through exchange interaction (usually known as Kondo effect),[30–36] and constructive quantum interference (known as weak localization (WL)).[25–27] These three mechanisms usually manifest themselves as well in an upturn of resistivity at low temperature, which is observed in the thinner films of Cu/BHT ratio of 5 at temperatures below 20K. However, they have different temperature dependence and contribute to the resistance according to the following expression:[36,37]

$$R(T) = R_0 + R_{e-p}T^5 - R_{e-e}T^{\frac{1}{2}} - q\ln T + R_K(T) \qquad (1)$$

where $T$ is absolute temperature, $R_0$ includes temperature-independent contributions such as residual resistance due to disorder, and $R_{e-p}T^5$, $R_{e-e}T^{\frac{1}{2}}$, $q\ln T$, and $R_K(T)$ represent the contributions of electron-phonon coupling, EEI in disordered metals (Altshuler–Aronov model),[37,38] quantum interference, and Kondo-like effect, respectively. The Kondo-like contribution $R_K(T)$ can be further related to the Kondo temperature $T_K$ and a parameter $s$, that depends on the spin of the impurity, by an empirical relation:[36]

$$R_K(T) = R_{K0}\left[\left(\frac{T}{T_K}\right)^2 \left(2^{1/s} - 1\right) + 1\right]^{-s} \qquad (2)$$

where $R_{K0}$ is a prefactor. Since $T < 20$ K is sufficiently low to freeze out phonon motions and $\ln T$ fit gives significant deviation, while $T^{\frac{1}{2}}$ fit, and $R_K(T)$ fit by **Equation (2)** provide good fitting results (see **Table S1** for the values of fitting parameters), EEI and Kondo-like effect seem possible to be the mechanisms governing the scattering at low temperature. The curvature of the negative MRs is helpful to distinguish these two mechanisms. It is known that in the diffusive transport regime, EEI manifests as parabolic negative MRs,[28,29] which was not observed, as demonstrated by its non-parabolic relation with the magnetic field (**Figure S6**). In contrast, we found that the negative MRs at both 10 and 50 K can be perfectly fitted by the Khosla-Fisher model:[34,39]

$$[\rho(B) - \rho(0)]/\rho(0) = -b_1^2 \ln(1 + b_2^2 B^2) + b_3^2 B^2/(1 + b_4^2 B^2) \qquad (3)$$

which includes third-order terms in the exchange Hamiltonian and has been successfully explained the field dependence of negative MRs in various dilute magnetic conductors. Parameters $b_1$ and $b_2$ are related to various physical quantities associated with exchange interaction and $b_3$ and $b_4$ are related to the carrier mobility and the density of states.[34,39] This good fit, with all fitting parameters $b_1$– $b_4$ falling within the ranges similar to the literature[34,39] (**Table S2**), could indicate that the charge transport at low temperature is governed by Kondo-like effect, and not by EEI and weak localisation.[33,35–37][33,35] Further, in-depth study of these transport signatures is needed, but here we conclude that thin films of Cu/BHT ratio of 5 yield the most pronounced quantum coherent transport properties with spin-related scattering (without or without magnetic ordering) being directly observable, while films grown with longer reaction time and lower Cu/BHT ratio exhibit more classical, (near-)metallic transport signatures.

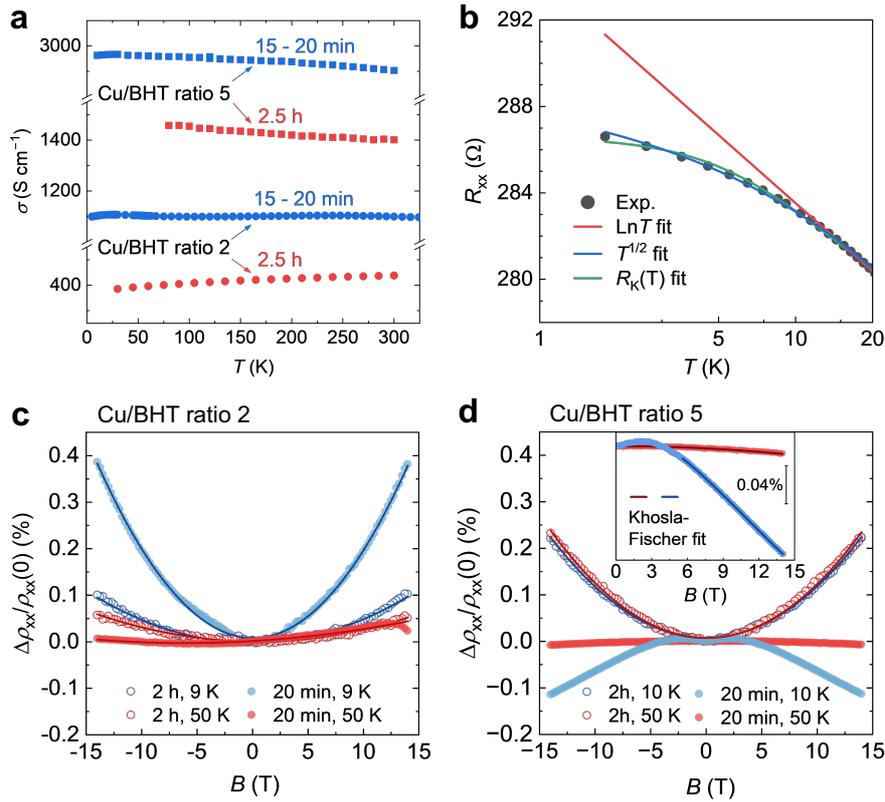

**Figure 5. Temperature-dependent electrical conductivity and magnetoresistance revealing the influences of reaction time on transport behaviors of Cu-BHT films of Cu/BHT ratios of 2 and 5.** **a**, Temperature-dependent electrical conductivities. **b,** Fittings of quantum interference (red line), EEI (blue line), and spin-dependent scattering (green line) on the experimental data. **c, d**, Field-dependent magnetoresistance at 9 – 10 K and 50 K for Cu/BHT ratios of 2 and 5 respectively. The solid lines in the main figures are parabolic fits, while those in the inset of figure 5d are Khosla-Fischer fits.

**Conclusions**

In this work, we first understood the growth evolution of Cu-BHT films in liquid-liquid interfacial synthesis; this allowed us to perform controlled synthesis and we obtained smoother and denser thinner films with a higher level of uniaxial face-on orientation, which were produced from the early stages of growth by shortening reaction time. Compared to the regular, thicker films synthesized by longer time, these thinner films generally have metallic charge transport with two- to three-fold higher electrical conductivities of up to > 3000 S cm$^{-1}$. We also uncovered low temperature transport properties exhibiting signatures of weak antilocalisation and Kondo-effect in these orientated thinner films. Our findings here not only demonstrate the possibility of obtaining oriented, high-performance 2D cCP films by a simple, fast and template-free route at a liquid-liquid interface, but also reveal the significance of microstructural control in observing quantum transport phenomena; these phenomena should be studied in more details to better understand the microscopic scattering mechanisms, the nature of the magnetic impurities and the potential role of spin textures in these 2D cCPs, in order to explore the potential of this class of novel 2D molecular materials in spin-electronic applications.

**Data availability**

All the data supporting the findings and mentioned in this study are included in the Article and its Supplementary Information files. Raw data are available from the corresponding authors upon request for academic and non-commercial purposes.

**References**

[1]   E. M. Johnson, S. Ilic, A. J. Morris, *ACS Cent. Sci.* **2021**, *7*, 445.

[2]   W. Jiang, X. Ni, F. Liu, *Acc. Chem. Res.* **2021**, *54*, 416.

[3]   M. Wang, R. Dong, X. Feng, *Chem. Soc. Rev.* **2021**, *50*, 2764.

[4]   R. Sakamoto, N. Fukui, H. Maeda, R. Toyoda, S. Takaishi, T. Tanabe, J. Komeda, P. Amo-Ochoa, F. Zamora, H. Nishihara, *Coord. Chem. Rev.* **2022**, *472*, 214787.

[5]   H. Maeda, K. Takada, N. Fukui, S. Nagashima, H. Nishihara, *Coord. Chem. Rev.* **2022**, *470*, 214693.

[6]   Y. Lu, P. Samorì, X. Feng, *Acc. Chem. Res.* **2024**, *57*, 1985.

[7]   S. Fu, J. Zhang, X. Li, E. Jin, L. Gao, R. Dong, Z. Wang, X. Feng, H. I. Wang, M. Bonn, *Nat. Rev. Mater.* **2025**, DOI 10.1038/s41578-025-00840-z.

[8]   Z. Wang, M. Wang, T. Heine, X. Feng, *Nat. Rev. Mater.* **2025**, *10*, 147.

[9]   Y. Misumi, A. Yamaguchi, Z. Zhang, T. Matsushita, N. Wada, M. Tsuchiizu, K. Awaga, *J. Am. Chem. Soc.* **2020**, *142*, 16513.

[10]  R. W. Day, D. K. Bediako, M. Rezaee, L. R. Parent, G. Skorupskii, M. Q. Arguilla, C. H. Hendon, I. Stassen, N. C. Gianneschi, P. Kim, M. Dincă, *ACS Cent. Sci.* **2019**, *5*, 1959.

[11]  J. Zhang, G. Zhou, H. I. Un, F. Zheng, K. Jastrzembski, M. Wang, Q. Guo, D. Mücke, H. Qi, Y. Lu, Z. Wang, Y. Liang, M. Löffler, U. Kaiser, T. Frauenheim, A. Mateo-Alonso, Z. Huang, H. Sirringhaus, X. Feng, R. Dong, *J. Am. Chem. Soc.* **2023**, *145*, 23630.

[12]  G. Skorupskii, K. N. Le, D. Leo Mesoza Cordova, L. Yang, T. Chen, C. H. Hendon, M. Q. Arguilla, M. Dinc, E. by Omar Yaghi, *Proceedings of the National Academy of Sciences* **2022**, *119*, e2205127119.

[13] X. Huang, P. Sheng, Z. Tu, F. Zhang, J. Wang, H. Geng, Y. Zou, C. A. Di, Y. Yi, Y. Sun, W. Xu, D. Zhu, *Nat. Commun.* **2015**, *6*, 7408.

[14] X. Huang, S. Zhang, L. Liu, L. Yu, G. Chen, W. Xu, D. Zhu, *Angewandte Chemie* **2018**, *130*, 152.

[15] T. Chen, J. H. Dou, L. Yang, C. Sun, N. J. Libretto, G. Skorupskii, J. T. Miller, M. Dincă, *J. Am. Chem. Soc.* **2020**, *142*, 12367.

[16] D. Sheberla, L. Sun, M. A. Blood-Forsythe, S. Er, C. R. Wade, C. K. Brozek, A. Aspuru-Guzik, M. Dincă, *J. Am. Chem. Soc.* **2014**, *136*, 8859.

[17] W. Zhong, T. Zhang, D. Chen, N. Su, G. Miao, J. Guo, L. Chen, Z. Wang, W. Wang, *Small* **2023**, *19*, 1.

[18] D. G. Ha, M. Rezaee, Y. Han, S. A. Siddiqui, R. W. Day, L. S. Xie, B. J. Modtland, D. A. Muller, J. Kong, P. Kim, M. Dincă, M. A. Baldo, *ACS Cent. Sci.* **2021**, *7*, 104.

[19] M. Choe, J. Y. Koo, I. Park, H. Ohtsu, J. H. Shim, H. C. Choi, S. S. Park, *J. Am. Chem. Soc.* **2022**, *144*, 16726.

[20] Z. Wang, L. S. Walter, M. Wang, P. S. Petkov, B. Liang, H. Qi, N. N. Nguyen, M. Hambsch, H. Zhong, M. Wang, S. Park, L. Renn, K. Watanabe, T. Taniguchi, S. C. B. Mannsfeld, T. Heine, U. Kaiser, S. Zhou, R. T. Weitz, X. Feng, R. Dong, *J. Am. Chem. Soc.* **2021**, *143*, 13624.

[21] S. Park, Z. Zhang, H. Qi, B. Liang, J. Mahmood, H. J. Noh, M. Hambsch, M. Wang, M. Wang, K. H. Ly, Z. Wang, I. M. Weidinger, S. Zhou, J. B. Baek, U. Kaiser, S. C. B. Mannsfeld, X. Feng, R. Dong, *ACS Mater. Lett.* **2022**, *4*, 1146.

[22] H.-I. Un, K. Iwanowski, J. Ferrer Orri, I. E. Jacobs, N. Fukui, D. Cornil, D. Beljonne, M. Simoncelli, H. Nishihara, H. Sirringhaus, *Nat. Commun.* **2025**, *16*, 6628.

[23] Z. Pan, X. Huang, Y. Fan, S. Wang, Y. Liu, X. Cong, T. Zhang, S. Qi, Y. Xing, Y. Q. Zheng, J. Li, X. Zhang, W. Xu, L. Sun, J. Wang, J. H. Dou, *Nature Communications* **2024**, *15*, DOI 10.1038/s41467-024-53786-1.

[24] H. Nakamura, D. Huang, J. Merz, E. Khalaf, P. Ostrovsky, A. Yaresko, D. Samal, H. Takagi, *Nat. Commun.* **2020**, *11*, 1.

[25] H.-Z. Lu, S.-Q. Shen, *Spintronics VII* **2014**, *9167*, 91672E.

[26] H. Wang, H. Liu, C. Z. Chang, H. Zuo, Y. Zhao, Y. Sun, Z. Xia, K. He, X. Ma, X. C. Xie, Q. K. Xue, J. Wang, *Sci. Rep.* **2014**, *4*, 1.

[27] M. Liu, J. Zhang, C. Z. Chang, Z. Zhang, X. Feng, K. Li, K. He, L. L. Wang, X. Chen, X. Dai, Z. Fang, Q. K. Xue, X. Ma, Y. Wang, *Phys. Rev. Lett.* **2012**, *108*, 1.

[28] J. Jobst, D. Waldmann, I. V. Gornyi, A. D. Mirlin, H. B. Weber, *Phys. Rev. Lett.* **2012**, *108*, 106601.

[29] T. Thornton, A. Matsumura ab, *Surf. Sci.* **1996**, *361*, 547.

[30] J. M. Salchegger, R. Adhikari, B. Faina, J. Pešić, A. Bonanni, *Phys. Rev. B* **2024**, *110*, 205403.

[31] S. Barua, M. C. Hatnean, M. R. Lees, G. Balakrishnan, *Sci. Rep.* **2017**, *7*, 10964.

[32] X. Y. Wang, M. H. Chen, C. X. Chen, Q. Zhao, F. H. Zhu, W. L. Yang, X. Y. Chen, M. L. Yan, R. F. Dou, C. M. Xiong, J. C. Nie, *Phys. Rev. B* **2025**, *111*, 094506.

[33] P. K. Leung, J. Slechta, J. G. Wright, *J. Phys. F: Metal Phys.* **1974**, *4*.

[34] R. P. Khosla, J. R. Fischer, *Phys. Rev. B* **1970**, *2*, 4084.

[35] J. Slechta, *J. Non. Cryst. Solids* **1975**, *18*, 137.

[36] X. Cai, J. Yue, P. Xu, B. Jalan, V. S. Pribiag, *Phys. Rev. B* **2021**, *103*, 115434.

[37] D. Nath, S. Chakravarty, U. P. Deshpade, A. V. T. Arasu, R. Baskaran, N. V. C. Shekar, *Current Applied Physics* **2022**, *34*, 122.

[38] J. Joseph, C. Bansal, K. J. Reddy, A. Rajanikanth, *AIP Adv.* **2020**, *10*, 125223.

[39] R. P. Khosla, J. R. Fischer, *Phys. Rev. B* **1972**, *6*, 4073.

[40] Z. Jiang, *J. Appl. Crystallogr.* **2015**, *48*, 917.

[41] J. Ferrer Orri, T. A. S. Doherty, D. Johnstone, S. M. Collins, H. Simons, P. A. Midgley, C. Ducati, S. D. Stranks, *Advanced Materials* **2022**, *34*, 2200383.

[42] E. O. D. N. Johnstone, P. Crout, S. Høgås, B. Martineau, J. Laulainen, S. Smeets, S. Collins, E. Jacobsen, J. Morzy, E. Prestat, T. Doherty, T. Ostasevicius, H. W. Ånes, T. Bergh, R. Tovey, "pyxem/pyxem: pyxem 0.12.0," can be found under https://doi.org/10.5281/zenodo.3968871, **n.d.**

[43] C. Ophus, S. E. Zeltmann, A. Bruefach, A. Rakowski, B. H. Savitzky, A. M. Minor, M. C. Scott, *Microscopy and Microanalysis* **2022**, *28*, 390.

## Acknowledgments

H.-I.U., H.S., K.M., and H.N. acknowledge the support from the Engineering and Physical Sciences Research Council (EPSRC) and the Japanese Society for the Promotion of Science (JSPS) through a core-to-core grant (EP/S030662/1). H.S. thanks the Royal Society for a Research Professorship (RP/R1/201082). I.J. acknowledges funding from a Royal Society University Research Fellowship (URF\R1\231287). J.F.O. acknowledges funding from EPSRC Nano Doctoral Training Centre (EP/L015978/1). SED studies were supported by the access to e02 at ePSIC Diamond Light Source (MG32017). We thank the Diamond Light Source Beamline I-07 for beamtime (SI35227, SI35227-1) and Jonathan Rawle for GIWAXS measurement assistance. H.-I.U. thank Deepak Venkateshvaran, who is a Royal Society University Research Fellow in the University of Cambridge, for reviewing the manuscript and providing useful feedback.

## Author contributions

H.-I.U. conceived the project, prepared all samples and devices, designed and conducted all experimental characterizations except for the SED characterization which was done and analysed by J.F.O independently under C.D. and S.D.S supervision. I.E.J. and H.-I.U performed the GIWAXS measurements and I.E.J. carried out the initial data processing of the GIWAXS raw data. N.F. synthesized the BHT molecule under H.N. supervision. H.-I.U analysed the data and led the scientific development of this work in H.S. group. H.S. provided resources for the project. H.-I.U. wrote the paper and revised it with contribution from H.S. All authors reviewed the manuscript.

## Competing interests

Authors declare that they have no competing interests.